# Molecular-Size Control of Properties of Therapeutic Nano-Paper Allows for Selective Drug Storage in Small Doses

Elisabeth Erbes[1,2,†], Naireeta Biswas[1,2,†], Calvin J. Gavilett[1,4], Matthias Schwartzkopf[1], Krishnayan Basuroy[1], Qing Chen[1], Andrei Chumakov[1], Susann Frenzke[1], Marc Gensch[1,5], Korneliya Goordeyeva[3], Patrycja Kielb[6,‡], Sonja Kirchner[6], Volker Körstgens[5], Peter Müller-Buschbaum[5,7], Henrike M. Müller-Werkmeister[6], Jan Rubeck[1], Sreevidya Thekku Veedu[1], Jose de Jesus Velazquez-Garcia[1], Vivian Waclawek[1], Daniel Söderberg[3], Stephan V. Roth[1,3,*], Simone Techert[1,2,*]

[1] Deutsches Elektronen-Synchrotron DESY, Notkestraße 85, 22607 Hamburg, Germany.

[2] Institute for X-ray Physics, Goettingen University, Friedrich Hund Platz 1, 37077 Goettingen, Germany.

[3] Department of Fibre and Polymer Technology, KTH Royal Institute of Technology, 100 44 Stockholm, Sweden.

[4] Department of Mechanics, KTH Royal Institute of Technology, 100 44 Stockholm, Sweden

[5] TUM School of Natural Sciences, Department of Physics, Technical University Munich, Lichtenbergstraße 1, 85748 Garching, Germany

[6] Institute of Chemistry – Physical Chemistry, Potsdam University, Karl-Liebknecht-Str. 24-25, 14476 Potsdam, Germany

‡ Current address: Clausius Institute of Physical and Theoretical Chemistry, University of Bonn, Wegelerstr. 12, 53115 Bonn, Germany

[7] Heinz Maier-Leibnitz Zentrum (MLZ), Technical University Munich, Lichtenbergstraße 1, 85748 Garching, Germany

* Correspondence to: simone.techert@desy.de, stephan.roth@desy.de, svroth@kth.se

† Equal contributions to this work

ABSTRACT

A novel concept of nano-scaled interwoven templates for drug delivery with alternating hydro- and lipophilicity properties is introduced. They are built from cellulose and peptide hydrogel in tandem, and characterized by a nano-stacked interwoven design, thus enabling for tuning the lipophilicity in the mesh nano-domains in which drug candidates of complementary lipophilicities can be embedded. This allows for low-dose-controlled consumption and therapeutic applications. Time-resolved and in-situ grazing incidence X-ray scattering studies confirm the design of the therapeutic nano-paper and create conditions suitable for the drug storage of complementary properties. The molecular design has the potential of a locally controlled, site-specific drug release on a beyond-nanomolar scale. Generalized, the design may contribute to facile developments of personalized medicine.

INTRODUCTION

Not only the worldwide COVID-19 disease as an example of infective pandemic, but also the increasing number of other illnesses difficult to treat (like some types of cancer) or individual medical care (like immunization or vaccination) demonstrate intensive research needed in individual therapeutic treatments or precision medicine *(1)*.

In the current proof-of-principle study, we have selected different types of drugs and pharmaceuticals, which have been developed for the treatment of cancer or viral infections *(2,3)*. These drugs have been discussed with different proposed inhibition mechanisms, with chemically complementary properties and even controversially debated in the literature *(4-11)*. The selection we made is rather based on the drug's chemical properties as "model systems" than based on their supposed antiviral, anti-cancer or other efficiencies. The results lead to a generalizable conclusion of tunable nano-dosing of drugs, which have the potential to be adapted to other types of drugs or therapies, in particular concerning vaccination into the dermal compartment *(12)*.

The impressive list and number of drugs systematically listed in publicly accessible data bases *(2, 3, 13)* (e.g. also been a source for the proposed anti-COVID drugs in the early stage of pandemic) give chemists a rich fund of molecular systems with even opposing chemical (hydrophilic and lipophilic) properties. While selecting the molecules with opposing chemical properties that also consist of a wide range of drug properties from the molecular databases, we asked the question whether it is possible to develop a nano-based drug matrix that may comprise drugs of opposed hydrophilic and lipophilic or in general complementary solubility properties. We also noted that most of the drug molecules studied imply a high risk of overdose, demanding very precise low-dose control. This has inspired us to develop a novel strategy for the uptake of drugs in nano-controllable doses for therapeutic purposes. The method makes it possible to administer drugs at controlled low doses and in a variable manner to reduce severe side effects.

For our study we have embedded various drug inhibitors in novel designed, nanometer-confined complementary drug delivery environments and matrices. Very recently it has been reported that functionalized nanopolyaniline/fibrin gel composite scaffolds may be used in the treatment of cardiovascular diseases *(14)*.

Our "therapeutic nano-paper" (TNP) consists of anti-infectious drugs, singular or in combination, embedded in hydrogels of tandem and nanoscale interwoven design of nanocellulose *(15,16)* and disordered nano-sized *(17)* peptide hydrogels *(18)* (Fig. 1). The novel design overcomes current restrictions of nanocellulose-based drug delivery templates *(19,20)*. Their structural and morphological integrity under performance conditions (Fig. 1A) have been studied with time-resolved *(19,20)* and *in-situ* grazing incidence small- and medium angle X-ray scattering (GISAXS and GIMAXS) techniques *(21-24)* (Fig. 1B and 1C) down to the molecular scale. The studies including AFM and static and quasi-type *in-situ* FTIR *(25)* allow revealing the different interactions and characterizing the complementary chemical mechanisms of the drug's integration into the TNP. The experiments confirm for different active ingredients reliable conditions of the TNP design and its creation suitable for drug storage. Thus the current study adds another important step into the development of studies on dermatological cellulose *(26-29)*. The template of the TNP consists of a mixture of hydrophilic carboxy-methylated cellulose (CMC) and peptide hydrogels (P). CMC is enzymatically active and does not form a hydrogel when dissolved in water. The CMC stacks alter with peptide hydrogels (P) consisting of variable hydrophilic (Threonin, Serin, Glutamin etc.) and hydrophobic (Glycine-Lysin-Phenylalanine-Glycine) (GLFG)-containing, long-chain amino acids allowing for tuning the hydrophilicity of the P template and therefore a selective implementation of drugs (for the full peptide sequence see appendix). CMC and its derivatives

*(17)* allow also for the systematic tuning of hydrogen bonds in the polysaccharide structures. This also enables a local uptake of the active ingredients in a specific chemical environment in CMC *(28)*. The hydrophobic character of polypeptide chains in P increases with the length of the GLFG peptide chains *(17)*. The long peptide chains arrange themselves into micellar structures by hydrophilic interactions and hydrophobic collapses *(18)* – similar to hydrophobic aerogels. Altering the GLFG peptide chains in size and composition allows for a controlled nano-tuning of the hydrophilic and hydrophobic domains.

RESULTS AND DISCUSSION

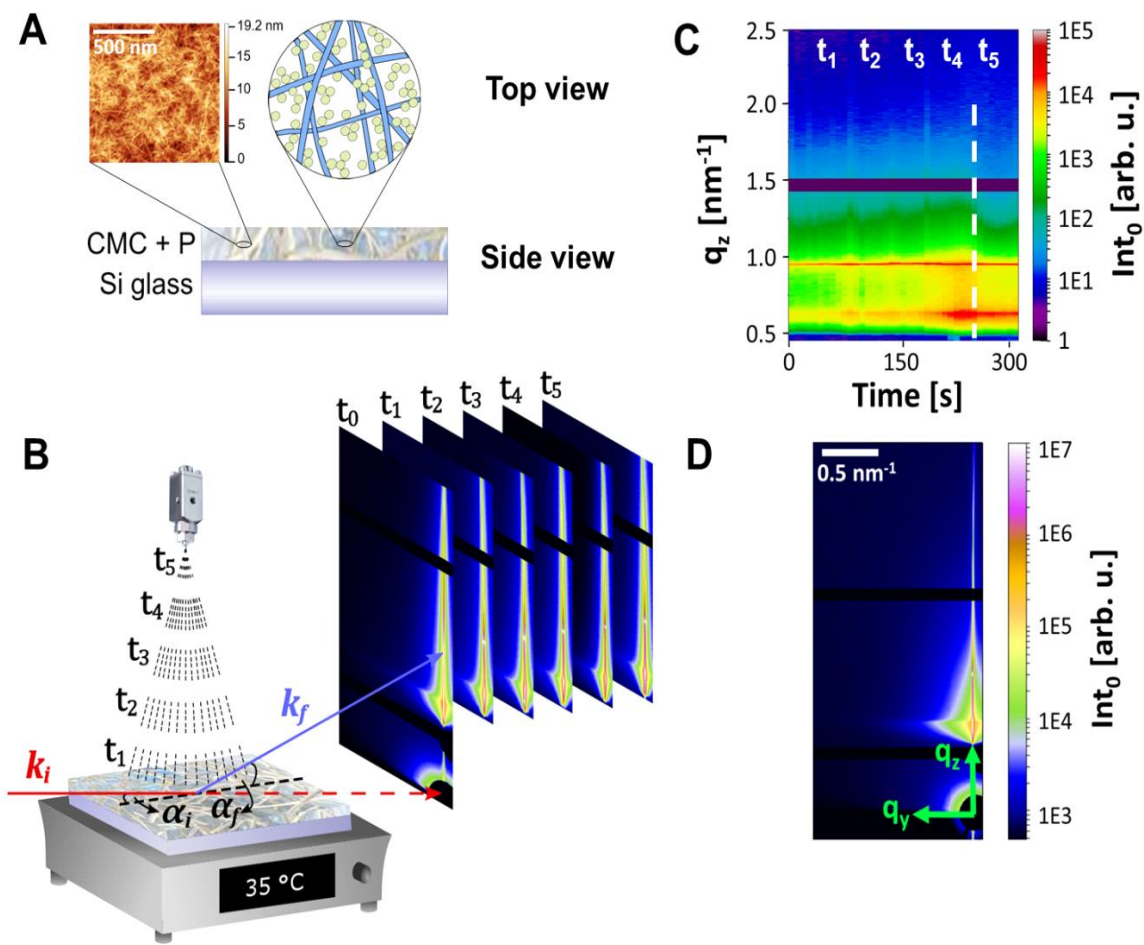

**Fig. 1**. **Time-resolved and *in-situ* grazing-incidence small-angle X-ray scattering (GISAXS) studies of the fabrication, properties, and functionality of "therapeutic nano-paper" (TNP)**. (A) Top view left: Atomic force microscopy (AFM) image of TNP CMC-layers. Top view right: interwoven cellulose (blue fibers, CMC) and peptide hydrogel layers (green points, P), allowing the nano-dosed inclusion of drugs (not shown here) with complementary properties. Side view: TNP template design with pre-mixed components. (B) Scheme of the spray set-up used at the P03 beamline at DESY. *In-situ* spray deposition of the drug's solution is performed while analyzing the layering process structures with GISAXS on-the-fly. $t_0$ denotes the time point before spraying drugs and, therefore, the scattering pattern of pure (CMC + P) TNP template without drugs. $t_1$ to $t_5$ are the *in-situ* GISAXS scattering pattern for every spraying cycle at time point $t_{j = 1,…,5}$. (C) Every *in-situ* spraying cycle is stroboscopically followed by a fast readout of the detector, which gives additional time-information between the spraying cycles. C shows the temporal evolution map of the stroboscopically followed density changes along $q_z$ in one *in-situ* experiment of five spraying cycles. The begin of the figure depicts $t_0$ before the spraying starts ($t_0 = 0$ s),

followed by the different spraying cycles. The dashed line depicts the last spraying cycle, and the end of the figure corresponds to the end of the drying process. (D) A typical 2-dimensional GISAXS pattern for the last time point $t_5$.

An overview of the GISAXS curves of selected drugs embedded in TNP is shown in Fig. 2A. Additional GISAXS analysis is found in Fig. S1 (see Supplementary Information, SI). In the entire X-ray scattering range we observe significant intensity modulations when adding the different drugs. The nano- and molecular scale morphology of the TNP nanostructures are determined from the *in-situ* GISAXS (0.02 nm$^{-1}$ < $q_y$ < 0.1 nm$^{-1}$) and GIMAXS (0.1 nm$^{-1}$ < $q_y$ < 1.2 nm$^{-1}$) measurements in a two-step analysis:

(i) *GISAXS refinement of the fiber-shaped aggregates* (0.02 nm$^{-1}$ < $q_y$ ≤ 0.1 nm$^{-1}$): The GISAXS cuts within 0.02 nm$^{-1}$ < $q_y$ < 0.1 nm$^{-1}$ (red and blue areas in Fig. 2A and Fig. 2B) have been adapted by modelling isotropically distributed cylinders / fibers with prolate radii dimensions *(23-27)*. The size fitting is limited by the resolution of the GISAXS data for $q_y$ ≤ 0.02 nm$^{-1}$. The structural parameters obtained from these adjustments have been chosen as limits for the next structural refinement steps.

(ii) *GIMAXS fractal refinement* (0.1 nm$^{-1}$ < $q_y$ < 1.2 nm$^{-1}$): Drug-nanobodies with electron density modulations of molecular orders in the range of 0.1 nm$^{-1}$ < $q_y$ < 1.2 nm$^{-1}$ (green area in Fig. 2A and Fig. 2B) have been incorporated into the drug templates in different concentrations. These intensity regions also show drug-dependent intensity modulations in the GIMAXS regime. We have evaluated them using fractal analysis schemes, as explained for Fig. 3 following the analysis outlined in *(31-34)*. For further information about the analysis see SI.

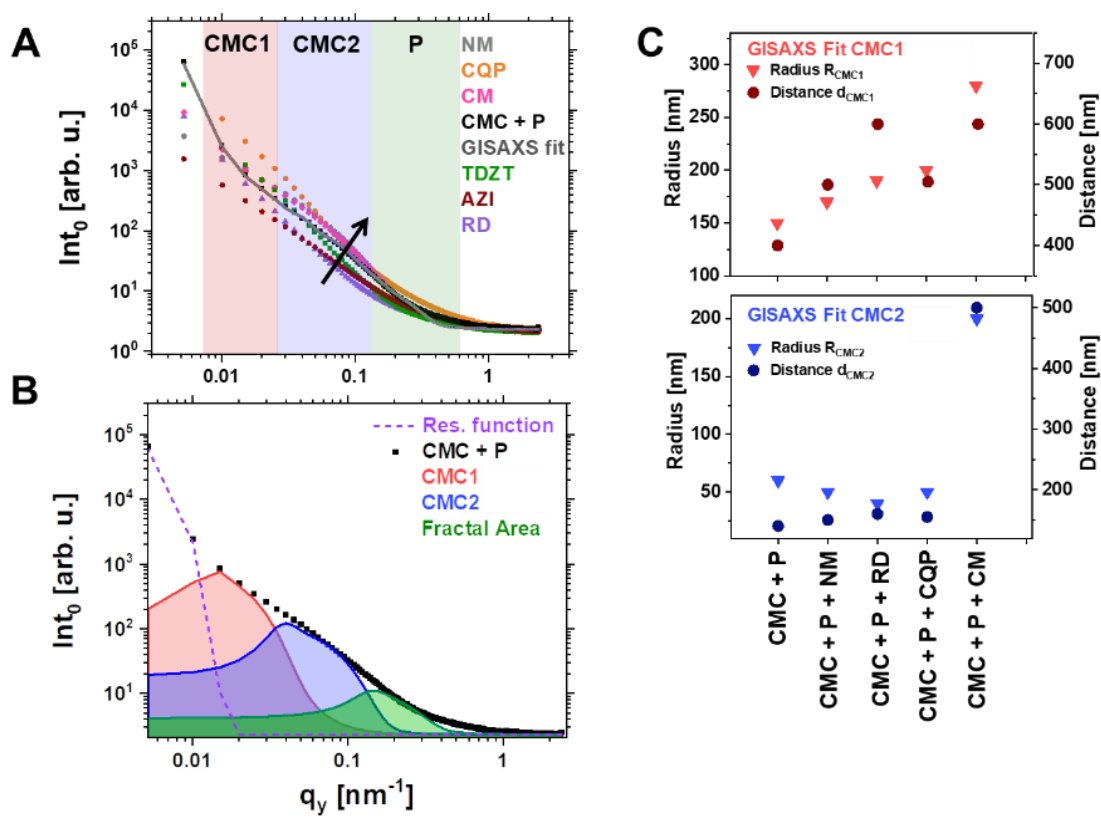

**Fig. 2**. **The different TNP nano-layers show characteristic domain formations in the different scattering regimes.** Domains with larger fiber-shaped CMC aggregates are highlighted as red, and domains with middle-sized CMC units highlighted as blue areas. The green area of long-chain peptide units P forming fractal networks (green area) are analyzed as follows: (A) Overview of all investigated drugs (for the abbreviation

see footnote). Depending on their chemical properties, they embed themselves in the different domain areas relative to the pure TNP template (black line). The arrow is added as an eye-guide. (B) GISAXS and GIMAXS curves of the pure TNP structures without drugs (CMC+P). The green curves indicate 10 nm size structures typical for CMC and starting to overlap with fractal structures formed by small (1-5 nm size) drug molecules. This region needs to be analyzed separately. (C) Top: refined fiber-shaped aggregation radii $R_{CMC1}$ and inter-fiber distances $d_{CMC1}$. Bottom: refined fiber-shaped aggregation radii $R_{CMC2}$ and $d_{CMC2}$ of the TNP and TNP loaded with drugs for the red and blue scattering areas (see also SI). The error bars lay within the size of the plotted data points.

For the pure TNP (CMC+P), we find high GISAXS intensity modulations within the red ("CMC1") and the blue ("CMC2") scattering regime (0.02 nm$^{-1}$ < $q_y$ ≤ 0.1 nm$^{-1}$ in Fig. 2A and Fig. 2B). Their analysis according to scheme *(i)* indicates the formation of one large ("CMC1") and one middle-sized ("CMC2") CMC structure with aggregate radii of $R_{CMC1}$ = 150 (± 10) nm and $R_{CMC2}$ = 50 (± 5) nm.

The same types of fibrillar structures (CMC1 and CMC2) are found when analyzing the GISAXS intensity modulations for pure CMC (see SI, Table S3). When the drugs are embedded in the TNP (Fig. 2C), the large fiber-shaped CMC aggregates CMC1 swell in size in the drug's order NM, RD, CQP and CM (Fig. 2C, top; for the abbreviations see foodnote). The CMC2 fibers do not change their size (Fig. 2C, bottom), except for CM loading which shows additional CMC aggregation behavior. We observe the same trends for the average distances between the fiber-shaped aggregates in the TNP, $d_{CMC1}$ and $d_{CMC2}$, and their polydispersity (SI, Table S3).

The big changes in the GISAXS and GIMAXS data prove a significant embedding of the drugs into the TNP and their strong influence on the TNP nanomorphology. Fig. 3A-F shows the refinement of the *(i)* GISAXS and *(ii)* GIMAXS scattering curves for the drugs CM (Fig. 3A, B), CQP (Fig. 3C, D) and RD (Fig. 3E, F). The representative molecular shapes of the drugs are a prolate ellipsoid with two active side group at both ends for CM, an almost flattened, oblate disk with one active side group terminus for CQP, and a sphere extending with two active side group arms each into all three dimensions for RD.

Continuing this molecular-shape description, the selected drugs characteristically form tighter or looser hydrogen bonding networks with the TNP environment. This depends on the position and type of the chemical side groups relative to the shape of the drug (Fig. 3G and SI, Table S4) and thus according to their (hydrophilicity or) lipophilicity character: Pure P-networks in CMC as reference are characterized by a high nano-porosity, whose electron density structures modulate on the molecular length scale with a peptide radius of $R_P$ = 0.471 ± 0.001 nm, and one order of magnitude larger inter-peptide distances of 5 ± 0.5 nm (SI, Table S3, particle center to center distance EIAS). In Fig. 3G, the fractal radius $R_P$ defines the smallest self-reproducing structural unit within the network scaled in molecular dimensions. The fractal radius $R_P$ increases with increasing drug's sizes from CM via CQP to RD.

Particularly, since CM preferentially adheres to the fibril-shaped CMC aggregates (blue and red area Fig. 2A – B, Fig. 2C), the CMC+P+CM network's fractal radius in Fig. 3G (bottom light green, left side and SI) equals with 0.45 ± 0.08 nm the radius of isolated CM molecular units (0.55 nm, determined through quantum chemical simulations (DFT), see SI), which are slightly larger than the smallest P unit (0.47 nm) and can thus be distinguished from them (Fig. 3A, B). On the contrary, the correlation length $\xi_P$ in Fig. 3G (right side) refers to the fractal domains' cluster size and shows a reversed order when correlating to the drug's sizes: decreasing from RD to CM to CQP.

We gained additional structural information by adding FTIR studies of the TNP components. $D_2O$ was used as solvent. Fig. 3H (and SI) show the FTIR spectra of the amide I and amide II regions corresponding to signals of the peptide bonds (bands between 1640 – 1670 cm$^{-1}$ and 1420 – 1460 cm$^{-1}$ of the pure P-network (green) and loaded with CM (pink). The amide I band, centered at 1640 – 1665 cm$^{-1}$ exhibits a clear signature for beta-sheets and -turns (shoulder at 1669 cm$^{-1}$) in combination with presence of random structures (peak at 1648 cm$^{-1}$) *(27)*. When loaded with a drug (equal preparation conditions as in the GISAXS/GIMAXS studies), the amide I and amide II band with their typical side shoulders remain nearly unaffected suggesting that the beta-sheet-structures remain unaltered during drug loading of the TNP.

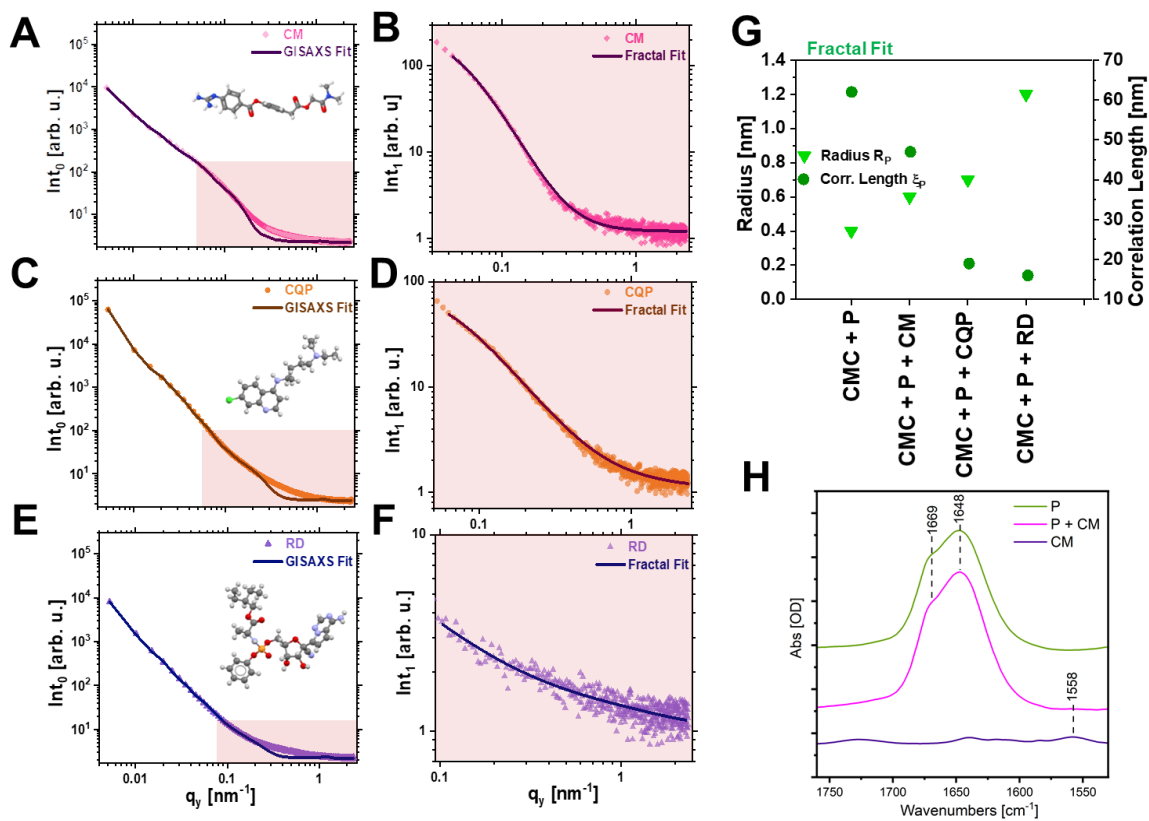

**Fig. 3. Representation of the structural adaptation of TNP for three representative drugs of this study: Camostat Mesylate (CM), Chloroquine-Phosphate (CQP), and Remdesevir (RD):** (A) GISAXS refinement of the CMC network for embedded CM, (B) GIMAXS refinement of the fractal P-network for the embedded CM drug; (C) GISAXS refinement of the CMC network for embedded CQP (it additionally shows a substructure formation at $q_y = 0.01$ nm$^{-1}$); (D) GIMAXS refinement of the fractal P-network for the embedded CQP drug; (E) GISAXS refinement of the CMC network for embedded RD; (F) GIMAXS refinement of the fractal P-network for the embedded RM drug. Due to code restriction, Int$_0$ is the scattering intensity w/o background subtraction, and Int$_1$ is the scattering intensity with subtraction of the GISAXS fit using three structures (see Fig. 2A, B). (G) The fractal radius for the P-network enriched by the drugs increases in the order CMC+P, CMC+P+CM, CMC+P+CQP, CMC+P+RD. The correlation length decreases in the order CMC+P. CMC+RD, CMC+P+CM and CMC+P+CQP. The errors are within the data points. (H) FTIR spectra of pure peptide (P, green), pure CM (CM, violet) and peptide with CM (P + CM, pink) in D$_2$O confirm a stability of the hydrogen bonding network: A small fraction of ordered β-turn formation (assigned based on amide I band between 1640 – 1670 cm$^{-1}$, green) as a secondary structure element is typical for the nanostructure of the P-network within the TNP and remains nearly unchanged when CM is added (pink spectra). The molecular radii of gyrations have been determined from quantum chemical gas phase calculations (DFT), the structures are shown in the insets of A), C) E).

What does this observation now implies including the drug's pharmaceutical properties like their *lipophilicity or hydrophilicity*? For getting a more detailed understanding of the drugs solubility in the TNP, we correlate our structural results with the so-called *lipophilicity or hydrophilicity* parameters of the drugs: According to their definition *(32-43)*, lipophilicity or the lipophilicity parameters of drugs are correlated to their solvation entropies driving hydrophilic and hydrophobic effects of drugs when dissolved or embedded into pharmaceutical relevant matrices like, in our case, the novel TNPs. As already mentioned in the explanation of Fig. 3, the majority of known drugs contains ionizable groups which are charged at physiological pH. For considering the pH-corrected properties, high performance liquid chromatography (HPLC)-based logD$_{7.4}$ values had been introduced. LogD$_{7.4}$ scales the lipophilicity to octanol/water (pH 7.4) standards and is reported in various data banks *(37)*. Negative logD$_{7.4}$ values (logD$_{7.4}$ <0) indicate high hydrophilicity, and positive logD$_{7.4}$ values (logD$_{7.4}$ >0) indicate low hydrophilicity and high lipophilicity.

Another measure for the hydro- or lipophilicity of the drugs is the protein binding value in % *(38)*. Values < 50% indicate high hydrophilicity, values > 50% high lipophilicity of the drugs. In Fig. 4 we correlated the derived structural parameters of the drug-loaded TNPs with these drugs characterizing pharmaceutical (solubility) parameters.

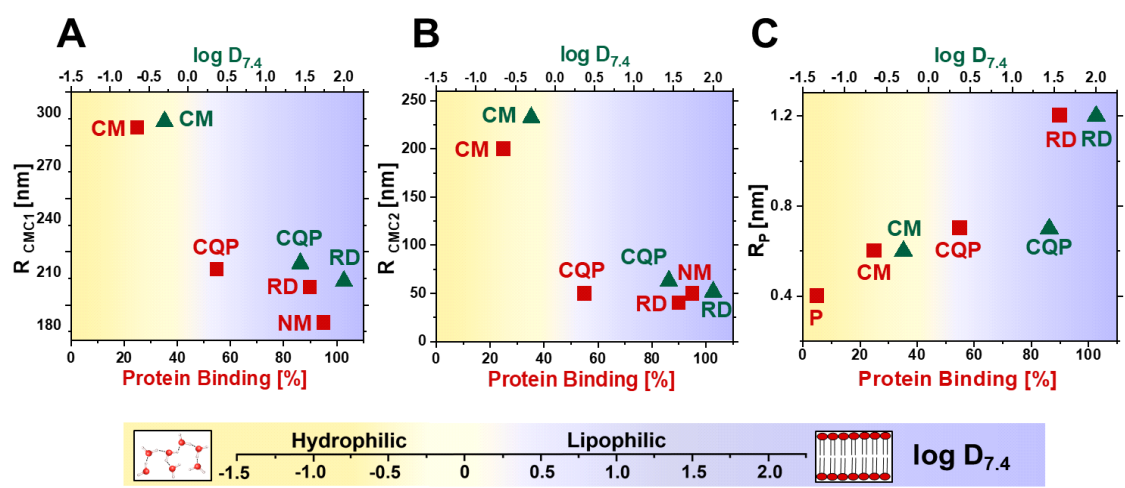

**Fig. 4. Correlation between the structural analysis of the drug-loaded TNP and the lipophilicity parameter of the drugs.** Top: Dependencies on the logD$_{7.4}$ value; Bottom: dependencies on the protein binding [%] values, both tabulated solubility characteristics of pharmaceuticals. (A) CMC1 large fiber R$_{CMC1}$ of CM, CQP, RD and NM. (B) medium-sized fiber R$_{CMC2}$ (50 nm) of CM, CQP, RD and other studied drugs and (C) fractal radii of the CMC and P-domains (R$_P$) of the TNP of the different drugs as a function of the lipophilicity parameters logD$_{7.4}$ (top) and %-protein binding values (bottom).

We can now give a first concise picture of the various static hierarchies of the TNP structures and how they are changed when loaded with drugs of oppose hydro- to lipophilicity in a static description (Fig. 2C top, Table S3 and S4, SI, Fig. 4). The larger radii of gyration of the CMC1 fiber-like subphase (aggregates), R$_{CMC1}$, increase characteristically without and with peptide and drug loading from R$_{CMC1}$(CMC+P) = 150 ± 12 nm to R$_{CMC1}$(CMC+P+RD) = 190 ± 20 nm, to R$_{CMC1}$(MC+P+CQP) = 200 ± 15 nm, and to R$_{CMC1}$(CMC+P+CM) = 280 ± 25 nm.

Quantitative values for the lipo-/hydrophilicity of the drugs are given in terms of logD$_{7.4}$ values. As the drugs become more lipophilic, the logD$_{7.4}$ values increase from negative to positive values, resulting in tighter packing of CMC1-TNP and a decrease in R$_{CMC1}$ as a function of logD$_{7.4}$ or protein binding (Fig. 4A). The system segregates during TNP formation, leading to a strong separation of the CMC and P domains, allowing the structures to organize themselves more densely. CM exhibits the highest hydrophilicity among the drugs studied and causes the most significant swelling of the CMC1 network. This is evidenced by the substantial increase in R$_{CMC1}$ from 190 ± 12 nm to 280 ± 25 nm. In contrast, CQP and RD, which are relatively lipophilic, show less interaction with the CMC network, resulting in R$_{CMC1}$ values of around 190-200 nm.

R$_{CMC2}$ is an intermediate structure of a CMC phase consisting of smaller structures. The size of R$_{CMC2}$ for CMC, CMC+P, CMC+P+RD, and CMC+P+CQP ranges from 40-60 nm (Fig. 2C bottom, Table S3, SI, Fig. 4B). For CMC+P+CM, however, R$_{CMC2}$ is 200 ±18 nm, indicating a significant swelling characteristic for the CMC domain due to specific interactions with CM. The observed slight changes in R$_{CMC2}$ are consistent with the results obtained from R$_{CMC1}$, as shown in Figure 4B. The correlation between $R_{CMC2}$ decrease and log $D_{7.4}$ increase is comparable and evident.

The distance between the centers of particles CMC1 and CMC2 decreases as the fiber wall-to-wall distance decreases. The third structure, which yields intensity in the green region of Fig. 2B, ranges between 10 and 15 nm for CMC, CMC+P, CMC+P+RD, CMC+P+CQP (Table S3, SI), while being 20 ± 3 nm for CMC+P+CM. This indicates a clear segregation between the CMC and P domains, which are both increasing in size (not included in Fig. 4).

An inverse dependence of the smallest refined radii of gyration $R_P$ is observed when comparing the classical phase shape evaluation with the fractal analysis in Fig. 4C. These radii describe the fractal network of the peptide mesh/subphase. They also dependent on the lipophilicity parameter $logD_{7.4}$, but inverse to $R_{CMC1}$ or $R_{CMC2}$. Upon loading with drugs, the P-mesh undergoes a significant swelling from $R_P(CMC+P) = 0.471 \pm 0.001$ nm to $R_P(CMC+P+RD) = 1.22 \pm 0.06$ nm. $R_P$ gradually increase from $R_P(CMC+P) = 0.471 \pm 0.001$ nm to (CMC+P+CM) = 0.58 ± 0.01 nm to $R_P(CMC+P+CQP) = 0.74 \pm 0.05$ nm to $R_P(CMC+P+RD) = 1.22 \pm 0.06$ nm as $logD_{7.4}$ and the lipophilicity increases (Table S4, SI and Fig. 4C). Compared to unloaded TNP, the lipophilic P-phase of TNP slightly de-packs and structurally disorders, when loaded with more hydrophilic CM, yielding a very small $R_P$ increase (Fig. 4C). Dissolving hydrophilic CM in the hydrophobic P-phase does not lead to molecular solvation stabilization but it results in better solvation into the hydrophilic CMC1 and CMC2 subphases (increase of their radii of gyration, Fig. 4A / 4B), with a greater segregation of the CMC from the P phases: the hydrophilicity of the drug embedded significantly affects the extent of segregation of the subphases during TNP formation.

Figure 4 clearly demonstrates that CM is the most hydrophilic drug among those studied. As a result, it causes the most significant swelling of the hydrophilic CMC phases, as evidenced by the increase in $R_{CMC1}$ and $R_{CMC2}$. The addition of drugs to the fractal template leads to an increased $R_{CMC1}$, which in turn causes a decrease in fiber wall-to-wall distances and a reduction in correlation lengths $\xi_P$ between two CMC fibers. Table S3, SI demonstrates that $\xi_P$ decreases in the order CQP, CM, RD. This indicates that the degree of cooperative structural organization within the TNP becomes more pronounced as the drugs become more hydrophilic.

Lipophilic drugs, like CQP and RD, enter the peptide network instead of the porous cellulose network: as the CMC1 and CMC2 radii of gyration decreases with loading TNP with CQP and RD, the radii of gyration of that P-mesh ($R_P$) increases due to a preferred solvation within the lipophilic P-phase.

Concerning dynamic studies, we performed additional *in-situ* real-time GISAXS and GIMAXS studies of the TNP formation. Fig. 5A-C (and corresponding Fig. S3, SI) confirms drastic differences in the kinetics of the TNP production for the various drugs – in particular when reaching the drug's concentration saturation conditions at the end of the spraying process. The separate time points are defined by a spraying and drying cycles of one TNP layer (and between two spray pulses). For the analysis, we focus on the normalized intensity of CMC2 and P region in Fig. 2A, B (see also Fig. S3, SI) in order to differentiate between the dynamics of formation and restructuring of the cellulose and peptide domains.

The lipophilicity of the drug has a significant impact on the formation of TNP. However, the kinetic response function of the fractal network is more complex during TNP formation. The temporal analysis of the drug's interaction with TNPs, as depicted in the blue areas of Fig. 5A-C, was obtained from the normalized intensities in the blue areas of Fig. 2 (Fig. S3, SI). The in-situ experiment primarily monitors time-dependent changes in the middle-sized structural CMC2-fiber network. The green line depicts the time-evolution of the normalized intensities derived from the green area, which belongs to the fractally organized P-network of the TNP.

The original CMC2-fiber network partially disappears in all three systems as the blue curves decrease in intensity.

The rate constants are globally determined by the TNP layer-to-layer growth, as shown in Fig. 5 (and Fig. S3, SI). The half-time rate constants $k_{1/2}$ for the different systems, CMC+P+CM, CMC+P+CQP and CMC+P+RD are in general of the same order of magnitude and depend on the spraying doses/drying cycles. In concrete terms we find for the kinetics of the TNP formation with CM $\tau_{1/2}$(CMC+P+CM, P) = 35 ± 2 s, $\tau_{1/2}$(CMC+P+CM, CMC) = 55 ± 5 s, for the ones with CQP $\tau_{1/2}$(CMC+P+CQP, P) = 30 ± 2 s, $\tau_{1/2}$ (CMC+P+CQP, CMC) = 62 ± 5 s, and a rate reversing and fastening of the process for RD with $\tau_{1/2}$(CMC+P+RD, P) = 55 ± 5 s and $\tau_{1/2}$ (CMC+P+RD, CMC) = 20 ± 2 s.

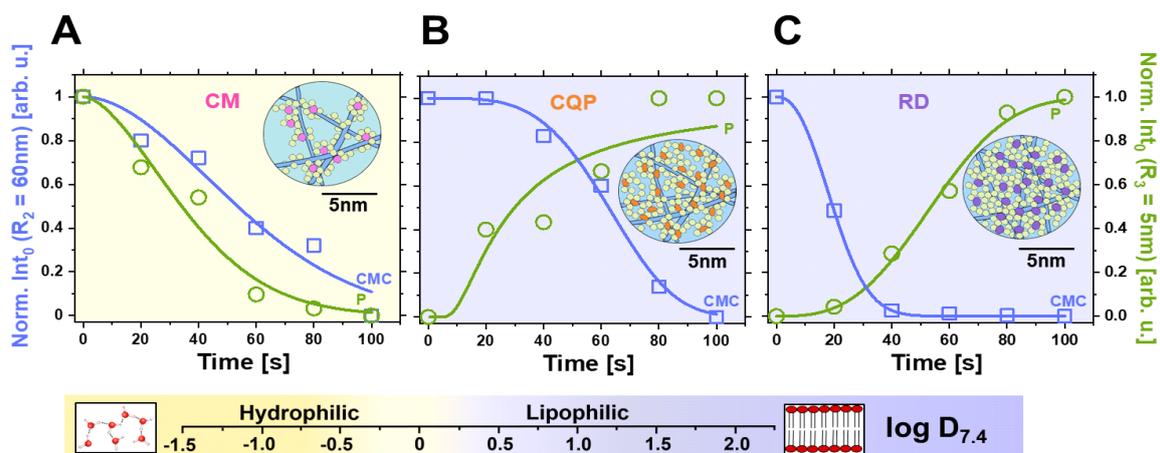

**Fig. 5. *In-situ* kinetic and structural analysis of the drug embedding during the spray production of the TNP.** All three TNP formations are kinetically driven. The insets present the final TNP-drug structures and morphologies after the last spraying cycle with CMC structures in blue, smaller dots for peptides and larger dots for the drugs. (A) For CM, a decomposition of the 50nm-middle-sized, fiber-shaped CMC aggregates (blue curve) and fractal P-network (green curve intensities) is seen: Both curves decrease. This implies a formation of specific P-drug fractal substructures around swollen fiber-shaped CMC aggregates. (B) For CQP, a dissolution of the 50nm-middle-sized CMC fibers is seen (blue curve decreases). Formation of aggregates within the P-network between the long fiber-shaped CMC aggregates is observed (green curve increases). (C) For RD, a faster dissolution of the middle-sized, 50 nm fiber-shaped CMC aggregates (blue curve decreases) under comparable concentration conditions is observed. The formation of the CMC-P-network with RD embedded in the fractal P-network and between the long fiber-shaped CMC aggregates is delayed and kinetically hindered (green curve increase).

Figure 5 clearly outlines the differences in the slope of the TNP growing rate constants between CM, CQP, and RD. It is observed that the hydrophilic CM system exhibits opposite kinetic behavior compared to the more lipophilic CQP and RD systems. The observed kinetic behavior in Fig. 5A for CM can be attributed to the structural decomposition of the original middle-sized, fiber-shaped CMC aggregates (blue curve decreases, left scale) and fractal P-network aggregates (green curve intensities decrease, right scale). The segregation process results in the formation of slightly disordered but specific CM-drug fractal-P-substructures around swollen fiber-shaped CMC aggregates, as demonstrated in Figs. 2-4. The FTIR results confirm that the β-sheet type of the P-TNP-structure persists during this reorganization in the nanodomains. Additionally, more lipophilic CQP still effectively dissolves the middle-sized CMC fibers (as evidenced by the blue curve decrease) but also forms ordered substructures and aggregates within the lipophilic P-network between the long fiber-shaped CMC aggregates (as evidenced by the green curve increase). These findings demonstrate the efficacy of CQP in restructuring the P- and the CMC-phases (aggregates). RD exhibits a similar kinetic behaviour to CQP (Fig. 5C). However, under comparable conditions, RD dissolves the middle-sized fiber-shaped CMC aggregates faster and more effectively than CQP. The formation of the CMC-P-network, with RD embedded in the fractal P-network and between the long fiber-shaped CMC aggregates, is delayed and kinetically hindered (as shown by the delayed increase of the green curve in Fig. 5C).

The original CMC2-fibril network partially disappears in all three systems as the blue curves decrease in intensity (Fig. S3, SI). However, the different drugs partially exhibit self-amplifying and autocatalytic reactions of increasing intensity, with comparable concentrations and slope characteristics but different time constants:

- The disassembly of the hydrophilic CMC2 fiber network in the hydrophilic CM's case results in the thinning out of the fractal hydrophobic P-network at comparable kinetic rates to the dilution of fiber-shaped CMC2-aggregates (as shown in Fig. 5A). It is worth noting that the disassembly of the hydrophobic P-network is autocatalytically accelerated compared to the ones of the CMC phases. Out of all the drugs, the most significant correlation length is demonstrated for CM (Fig. 3G). It is observed that CM forms clusters within the dissolved P- and CMC-mesh. The CMC1-aggregates are likely to accumulate on the surface of the large fiber-shaped structures, as suggested by the evidence (Fig. 2C top). The mechanism of this process is illustrated in the inset of Fig. 5A. This relocation of the P-assembled structures to other areas within the TNP implies that the disordered arms of the P-mesh loosen their network between the fibres and compactly arrange themselves with the drugs around the fibres. This relocation of the P-assembled structures to other areas within the TNP implies that the disordered arms of the P-mesh loosen their network between the fibres and compactly arrange themselves with the drugs around the fibres. These findings demonstrate a clear understanding of the incorporation of CM into the TNP.
- More lipophilic CQP and RD dissolve CMC2-aggregates of fiber structures in classical 1st order reactions, with RD being faster than CQP. The kinetic analysis clearly demonstrates that CQP autocatalytically enriches the interstitial P-mesh, thereby accelerating the increase in TNP structural compactness, outpacing the dissolution of the fiber-shaped CMC2 aggregates (as shown in Fig. 5B, inset). Both prefer solvation within the P-mesh.
- Lipophilic RD enhances the P-phase formation, as demonstrated by the intensified green curve in Fig. 5C. The time constant of this process follows a classical first-order kinetic law, similar to the disassembly of CMC2-aggregates, rather than an autocatalytic kinetics.

The CMC2 aggregate domains for all three drugs consist of two sub-phases: one forming larger or more swollen CMC fibers (Fig. 2C), and one with much smaller fiber aggregation radii, slightly larger than the typical fractal network dimensions (SI, Table S3, S4). Our *in-situ* and real-time measurements of TNP production demonstrated two important aspects for storing these drugs/drug models with opposing lipophilicity.

- As to Fig. 5, the nanostructure of the drug matrix is crucial in kinetically controlled drug delivery, in particular for low-dosage treatments. Incorporating hydrophobic (cellulose) and hydrophilic (peptides) nano-units can create optimized direct next-neighbor conditions for different drugs.
- Regulating the local concentration of matrix/host formation during TNP fabrication can be achieved by controlling phase transformations and segregation when drugs are embedded. Among the studied drug systems, it is evident that CM undergoes the most significant structural rearrangement when embedded into the TNP. This is because the structural response function of CM adapts to its compartment environment and adjusts the hydrogen network to backbone structures such as the peptides. The drug/host system of CM embedded in our TNP (i.e. CMC + P) is highly promising due to the structure of CM and the flexibility of the peptide network.

CONCLUSION

In summary, the autocatalytic embedding of CM in TNP is unequivocally supported by CM's specific chemical interactions with the P-network/mesh. These interactions lead to a preferential restructuring around the fibril-shaped CMC aggregates. Out of all the drugs and model drugs studied, CM displays the most exceptional matrix solubility behavior due to its remarkably high degree of hydrophilicity. CQP exhibits intensive chemical interactions with the P-network between the fiber-shaped CMC-aggregates. RD selectively enriches within the hydrophobic P-network intermixing with the CMC-phases, inducing structural disorder properties in the CMC network during embedding. CM exhibits hydrophilic properties, CQP is weakly lipophilic, while RD is highly lipophilic.

Structurally, the hydrophilic interactions of CM lead to two autocatalytically enhanced reorganizations of both the fractal P and the middle-sized CMC2 nano-networks from their original shapes towards a higher-dimensional network. During dissolving CM in the CMC network, the process of diluting the CMC2 nano-cylinders causes a significant swelling in size of the larger CMC1 nano-cylinders. We interpret this as a reformation of the TNP units around the CMC1 nano-cylinder network, at the expense of the CMC2 and in particular P-units, in favor of the CMC1 units.

In contrast, the highly lipophilic RD strongly dilutes within the hydrophobic P-network forming solvated structures and effectively increasing its fractal radius due to the enlargement of the structural P-solvation shell. The slightly higher density of the CMC-network is also a direct result of this. The CMC2 nano-cylinders undergo partial dissolution, while the CMC1 nano-cylinders experience slight swelling. It is important to note that CQP, which has weak lipophilicity, shows a similar trend though more pronounced.

TNP is an ideal carrier platform which can be customized for all type of drugs, whether hydrophilic or lipophilic. This is because TNP can individually reorganize around the drug, being hydro- or being lipophilic resulting in an drug-specific stabilization of the drug's solvation by structurally patterning the overall hydro- or lipophilicity of the TNP. Predictively, hydrophilic drugs will autocatalytically reshape the TNP towards a reorganization stabilizing the drug's solvation by remaining TNP's overall lipophilicity. This makes TNP to an excellent carrier material for all types of drugs – hydrophilic as well as lipophilic ones.

**Acknowledgments:**

**Funding:** The project is supported by HGF-Recruitment-DESY, HGF-InnovPool-FISCOV and HGF-POF-IV-RT3. ST and HWM are in particular grateful to financial support of the ESP program of the CMWS/DESY. JVG is grateful to financial support through 217133147/SFB 1073 of the DFG. SR and ST acknowledge DESY and DESY-FS for beamtime granting and special access during a CORONA-call of PETRA III. CB and SR acknowledge funding from DESY strategic fund (DSF) "Investigation of processes for spraying and spray-coating of hybrid cellulose-based nanostructures". This work benefited from the use of the SasView application, originally developed under NSF award DMR-0520547. SasView contains code developed with funding from the European Union's Horizon 2020 research and innovation program under the SINE2020 project, grant agreement No 654000.

**Author contributions:** The therapeutic nano-paper and its chemical subunits have been designed by: EE, SF, CB, QC, KG, DS, SR (CMC); NB, KB, ST (peptide); STV, JVG, SF (drugs inclusion). *In-situ* thin films have been grown by EE, MS, CG and QC. Time-resolved and *in-situ* GISAXS has been collected by NB, EE, MS, JR, MG, CG, VW, AC, QC, SR and ST under special pandemic operation constrains. CG, EE and DS added AFM studies. The chi2 analysis has been performed by EE and NB, with contributions (MS) and helpful discussions from MS, VK and PMB; EE, NB and ST integrated additionally the SASView analysis (also in the SI). PK, SK and HMW contribute with the in-situ FTIR studies. NB, EE, SVR and ST wrote the manuscript.

**Acknowledgments: Funding:** The project is supported by HGF-Recruitment-DESY, HGF-InnovPool-FISCOV and HGF-POF-IV-RT3. ST and HWM are in particular grateful to financial support of the ESP program of the CMWS/DESY. JVG is grateful to financial support through 217133147/SFB 1073 of the DFG. SR and ST acknowledge DESY, a member of the Helmholtz association HGF, for granting beamtime at P03 beamline during a COVID19 fast

**Competing interests:** There are no competing interests.

**Data and materials availability:** Data about the experimental conditions, codes and materials are available in the main text or SI.